\documentclass[iop,revtex4,twocolumn,apj,numberedappendix,appendixfloats]{emulateapj}
\usepackage{apjfonts}
\usepackage[pagebackref=false,colorlinks=true,citecolor=blue,linkcolor=blue,breaklinks=true,bookmarks=true]{hyperref}
\usepackage{amsmath}
\usepackage{comment}

\newcommand{\msun}{$M_{\odot}$}
\newcommand{\msunyr}{$M_{\odot}~{\rm yr}^{-1}$}
\newcommand{\lsun}{$L_{\odot}$}
\newcommand{\ergs}{erg~s$^{-1}$}
\newcommand{\cmsq}{cm$^{-2}$}
\newcommand{\um}{$\mu$m}

\newcommand{\lx}{$L_{\rm X}$}
\newcommand{\lmir}{$L_{\rm 12\mu m}$}
\newcommand{\loiii}{$L_{\rm [OIII]}$}
\newcommand{\nod}{\nodata}
\newcommand{\chandra}{{\it Chandra}}
\newcommand{\xmm}{{\it XMM-Newton}}
\defcitealias{Fu15a}{Paper I} 
\defcitealias{Fu15b}{Paper II}

\begin{document}

\title{ X-ray Properties of Radio-Selected Dual Active Galactic Nuclei}

\author{
Arran C. Gross\altaffilmark{1}, 
Hai Fu\altaffilmark{1}, 
A. D. Myers\altaffilmark{2},
J. M. Wrobel\altaffilmark{3}, and
S. G. Djorgovski\altaffilmark{4}
}
\altaffiltext{1}{Department of Physics \& Astronomy, The University of Iowa, 203 Van Allen Hall, Iowa City, IA 52242}
\altaffiltext{2}{Department of Physics \& Astronomy, University of Wyoming, Laramie, WY 82071}
\altaffiltext{3}{National Radio Astronomy Observatory, P.O. Box O, 1003 Lopezville Road, Socorro, NM 87801}
\altaffiltext{4}{California Institute of Technology, 1200 E. California Blvd., Pasadena, CA 91125}

\begin{abstract}
Merger simulations predict that tidally induced gas inflows can trigger kpc-scale dual active galactic nuclei (dAGN) in heavily obscured environments. Previously with the Very Large Array, we have confirmed four dAGN with redshifts between $0.04 < z < 0.22$ and projected separations between 4.3 and 9.2 kpc in the SDSS Stripe 82 field. Here, we present \chandra\ X-ray observations that spatially resolve these dAGN and compare their multi-wavelength properties to those of single AGN from the literature. We detect X-ray emission from six of the individual merger components and obtain upper limits for the remaining two. Combined with previous radio and optical observations, we find that our dAGN have properties similar to nearby low-luminosity AGN, and they agree well with the black hole fundamental plane relation. There are three AGN-dominated X-ray sources, whose X-ray hardness-ratio derived column densities show that two are unobscured and one is obscured. The low obscured fraction suggests these dAGN are no more obscured than single AGN, in contrast to the predictions from simulations. These three sources show an apparent X-ray deficit compared to their mid-infrared continuum and optical [O\,{\sc iii}] line luminosities, suggesting higher levels of obscuration, in tension with the hardness-ratio derived column densities. Enhanced mid-infrared and [O\,{\sc iii}] luminosities from star formation may explain this deficit. There is ambiguity in the level of obscuration for the remaining five components since their hardness ratios may be affected by non-nuclear X-ray emissions, or are undetected altogether. They require further observations to be fully characterized.
\end{abstract}


\section{Introduction} \label{sec:intro}

In a universe dominated by cold dark matter and dark energy, galaxies are built up from a series of major and minor mergers. Cosmological hydrodynamical simulations have shown that galaxies like our Milky Way have acquired a large ex-situ component from a number of other galaxies during the cosmic history \citep[e.g., ][]{Rodriguez-Gomez16,Pillepich18}. Zooming in on the encounter, idealized merger simulations have predicted that gravitational torques from tidally-induced stellar bars can drive large amounts of interstellar gas to the central kiloparsec on short timescales. The sudden gas inflow could dilute gas metallicity in the interstellar medium, fuel nuclear starbursts, and even trigger supermassive black hole (SMBH) accretion in the central parsec \citep[e.g.,][]{Barnes91,Torrey12,Capelo17a}. As a result, simultaneous SMBH accretion could occur at the smallest, sub-galactic separations ($\lesssim$10\,kpc) when the tidal force is strongest \citep[e.g.,][]{Van-Wassenhove12,Blecha13,Rosas-Guevara18}. Mergers in this stage would appear as kpc-scale dual active galactic nuclei (dAGN) before nuclear coalescence. 

The first source to be observed with signs of kpc-scale dAGN activity is the radio galaxy 3C\,75 \citep{Owen85}, which shows a pair of radio jets originating from a pair of stellar nuclei separated by 7.7\,kpc. More recent case studies have identified a number of kpc-scale dAGN with optical and X-ray data, including LBQS\,0103$-$2753 \citep{Junkkarinen01}, NGC\,6240 \citep{Komossa03}, 3C\,294 \citep{Stockton04}, Arp\,299 \citep{Ballo04}, Mrk\,463 \citep{Bianchi08}, NGC\,3393 \citep{Fabbiano11}, and Mrk\,266 \citep{Mazzarella12}.
To assemble a statistical sample of dAGN with well-understood selection biases, various techniques have been developed to identify dAGN candidates and subsequently confirm them systematically. 
Effective candidate identification involves lower-resolution observations of larger samples: (1) double-peaked [O\,{\sc iii}] AGN in the DEEP2 Galaxy Redshift Survey \citep{Gerke07,Comerford09a} and in the Sloan Digital Sky Survey \citep[SDSS;][]{Wang09,Liu10a,Smith10}, (2) hard-X-ray-detected nearby AGN in the Swift Burst Alert Telescope (BAT) survey \citep{Koss10}, (3) radio AGN pairs in the VLA Stripe 82 survey \citep{Fu15a}, (4) mid-IR-color-selected AGN in WISE \citep{Satyapal14,Satyapal17}, (5) BPT-selected AGN in spectroscopic galaxy pairs in SDSS \citep{Ellison11,Liu12,Fu18}, and (6) the \chandra \ surveys of merging galaxies \citep{Brassington07,Teng12}.
Robust candidate confirmation requires high-resolution observations at various wavelengths, e.g., in optical and near-infrared (IR) with Keck adaptive optics imaging \citep{Fu11a,Rosario11}, \textit{HST} imaging \citep{Liu13,Comerford15,Liu17c}, and integral-field spectroscopy \citep{McGurk11,Fu12a}, in X-ray with \chandra\ \citep{Comerford11,Koss11,Koss12,Teng12,Liu13a,Comerford15,Ellison17} and NuStar \citep{Ptak15,Koss16}, and in radio with the Very Large Array \citep{Fu11b,Fu15b,Muller-Sanchez15} and the Very Long Baseline Array \citep{Tingay11,Deane14,Wrobel14,Bondi16,Liu17b}.
Identifying and confirming dAGN is a slow process; currently there are only a total of $\sim$30 dAGN with projected separations less than 10\,kpc in the literature \citep[see][for a compilation]{Satyapal17}. Despite the small sample, there is now strong evidence that AGN activities are highly correlated in mergers while the AGN duty cycle remains more or less unchanged \citep[e.g.,][]{Fu18}. 

Observations at X-ray wavelengths can provide an effective measure of the accretion power of SMBHs in AGN. Nuclear X-ray emission in the keV-band could be produced by several mechanisms related to accretion: (1) inverse-Compton scattered thermal emission from a hot ($\sim10^9$\,K) corona \citep{Liang77,Haardt93} embedded in a cooler ($\sim10^5$\,K) standard geometrically-thin optically-thick disk \citep{Shakura73}, (2) a hot ($\sim10^{12}$\,K at the Schwarzschild radius) advection-dominated accretion flow \citep[ADAF; see][for a review]{Yuan14}, and (3) synchrotron and self-Comptonized emission from radiative shocks at the base of relativistic jets \citep{Yuan02}. Although it is difficult to disentangle the contribution of the various mechanisms in individual AGN, generally speaking, the ``disk-corona'' model produces the ``big blue bump'' in the UV and soft X-ray and likely dominates at high Eddington ratios ($L_{\rm bol}/L_{\rm Edd} > 0.01$), while the ``big blue bump'' is absent in the ADAF and the ``disk-jet'' models which dominate at low Eddington ratios ($L_{\rm bol}/L_{\rm Edd} < 0.01$). 
At low X-ray luminosity ($\lesssim 10^{42}$\,\ergs), another complication arises:  extended kpc-scale emission may appear spatially {\it unresolved} at the \chandra\ resolution at $z \gtrsim 0.05$, where $1\arcsec\ \gtrsim 1$\,kpc. This makes it difficult to discern the X-ray emission due to the AGN central engine from emission of the surrounding host galaxy. Common nonnuclear X-ray emission from normal galaxies originates from (1) the hot interstellar medium (ISM) in hydrostatic equilibrium (often seen in early-type galaxies; e.g., \citealt{Forman85}) and (2) X-ray binaries, supernovae remnants, and hot winds produced by recent star formation (often seen in late-type galaxies and starburst galaxies; e.g., \citealt{Fabbiano89}). Such spatially extended sources of X-ray emission have typical luminosities between $10^{38}$ and $10^{42}$\,\ergs, approaching the level of nuclear X-ray luminosity in Seyferts ($10^{42} < L_{\rm 0.5-10 keV} < 10^{44}$\,\ergs; \citealt{Brusa07}). 

 Nuclear X-ray emission in the 0.5$-$10\,keV band can be effectively absorbed by Helium and heavier elements due to photoelectric absorption \citep{Morrison83}. The obscuration can be caused by a dusty molecular torus at $\sim$10-pc scales and the intervening ISM in the host galaxy. The former is suggested by the AGN unification scheme \citep{Urry95} and confirmed in observations of narrow-line FR\,II radio galaxies \citep[e.g.,][]{Sambruna99} and nearby Seyferts \citep[e.g.,][]{Garcia-Burillo16,Gallimore16,Alonso-Herrero18,Fabbiano18}. But the torus may be absent at low accretion rates, such as in the FR\,I radio galaxies in the 3CR sample \citep[e.g.,][]{Donato04,Balmaverde06,Evans06a} and the nearby low-luminosity AGN \citep[LLAGN with $L_{\rm 0.5-10 keV} < 10^{42}$\,\ergs; see][for a review]{Ho08}. In these cases, X-ray emission can only be obscured by the intervening ISM \citep[e.g.,][]{Gilli14}. Because the gas column density is elevated during a galaxy merger by tidally induced gas inflows, one would expect an excess of obscured AGN in late-stage mergers \citep[e.g.,][]{Hopkins05a}.

In this series, we identify dAGN candidates within the 92 deg$^{2}$ covered by the wide-area 2\arcsec-resolution VLA 1.4\,GHz survey of the Stripe 82 field \citep[][hereafter \citetalias{Fu15a}]{Fu15a}. Out of 17,969 discrete radio sources, we identify 52 candidate pairs, which show good positional alignments between the radio sources and their optical counterparts. Optical spectroscopy available for eight of these candidates at that time reveals six pairs with consistent redshifts. We then follow up on these candidates with 0.3\arcsec-resolution VLA 6\,GHz observations \citep[][hereafter \citetalias{Fu15b}]{Fu15b}. The higher-resolution radio imaging reveals two of these candidate pairs to be projections of jets from single sources. The remaining four pairs are confirmed as dual AGN with compact ($\lesssim$0.4\arcsec) nuclear radio emission and core luminosities between $37.3 < $ log$\,L_{\rm 5GHz}/{\rm erg~s}^{-1} < 39.4$. 

Here we present \chandra\ X-ray observations of this radio and optically confirmed sample of dual AGN. The X-ray data enable us to address two main questions: {\it Do dAGN follow the same multi-wavelength scaling relations as the general AGN population?} As well, {\it are dAGN obscured to a higher degree than the general AGN population?} This paper is organized as follows. In \S~\ref{sec:obs} we describe the \chandra\ observations and data reduction procedures. In \S\,\ref{sec:analysis} we present the X-ray photometry and spectral fitting procedures. We synthesize our previous observations and the deduced X-ray properties of the dAGN in \S~\ref{sec:results}, wherein we test whether the dAGN deviate from the scaling relations established by general AGN of similar luminosities and also address whether dAGN are potentially more obscured at X-ray wavelengths. We conclude with a summary of our results and discuss their implications in \S~\ref{sec:summary}. Throughout, we adopt a $\Lambda$CDM cosmology with $\Omega_{\rm m} = 0.3, \ \Omega_{\Lambda} = 0.7$, and $h$ = 0.7. As usual, the spectral index $\alpha$ and the photon index $\Gamma$ are defined such that $F_\nu \propto \nu^{-\alpha}$ and $n(E) \propto E^{-\Gamma}$.

\section{Chandra Observations} \label{sec:obs}

\begin{figure*}[!t]
\centering
\includegraphics[width=0.95\linewidth]{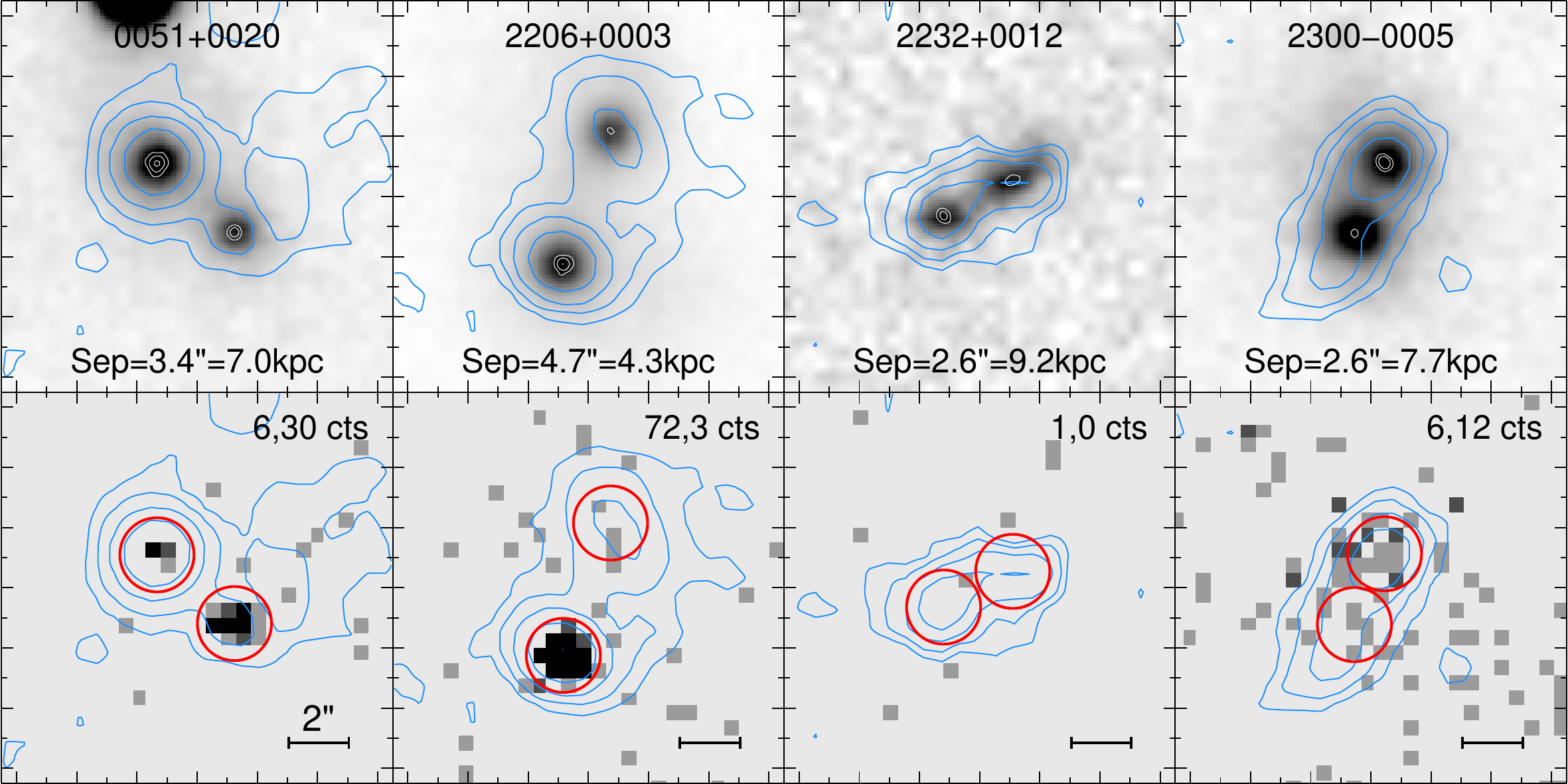}
\caption{Near-IR and X-ray images of the dual AGN, with radio contours. For each system, we show: ({\it top row}) a UKIDSS $H$-band image overlaid with the VLA 6\,GHz map ({\it white} contours; 0.3\arcsec\ beam) and the 1.4\,GHz map from the VLA-Stripe 82 survey ({\it blue} contours; 1.8\arcsec\ beam), with the source short designation, and the projected separation of the 6\,GHz radio cores; ({\it bottom row}) the ACIS image overlaid with the 1.4\,GHz map, X-ray photometry apertures ({\it red} circles), the full-band X-ray counts for the East ({\it left}) and the West ({\it right}) components, and a 2\arcsec\ scale bar. The X-ray images are displayed in their native $\sim$0.5\arcsec\ pixels. The 6\,GHz VLA contours are at ($+$3, $+$6, $+$24, $+$96)$\times$$\sigma$. The 1.4\,GHz VLA contours start at 2$\sigma$ and the levels increase exponentially to the peak S/N. Major tickmarks are spaced in  2\arcsec\ intervals. N is up and E is left for all panels. 
\label{fig:xrayimgs}}
\end{figure*}

Our sample consists of four radio-selected kpc-scale dAGN between $0.04 < z < 0.22$ in the SDSS Stripe 82 field  \citepalias{Fu15b}. All of the galactic nuclei show compact radio emission at a resolution of 0.3\arcsec. The radio cores show projected separations between 2.6\arcsec\ and 4.7\arcsec\ (or 4.3\,kpc and 9.2\,kpc). Under \chandra\ Cycle 18 Proposal \# 18700044, we observed the targets with the Advanced CCD Imaging Spectrometer (ACIS) in October 2016 (Obs ID: 19456) and September 2017 (Obs IDs: 19453-19455). The targets were placed near the aim point on the back-illuminated S3 chip. The average integration time is 24\,ks. 

Data reduction and analysis is carried out with the \chandra\ Interactive Analysis of Observations software \citep[CIAO v4.9;][]{Fruscione06}. Because the first observation and the last observation took place almost a year apart, we reprocess all of the data to the calibration database (CALDB v4.7.7) with \texttt{chandra\_repro}. We then focus on the data of the S3 chip (CCD ID: 7) in the calibrated energy range between 0.3 and 10\,keV by filtering the calibrated level 2 event file with \texttt{dmcopy}. Only the events with grades 0, 2, 3, 4, and 6 were included. We detect no significant flares in the background light curves. In the observation of 2206$+$0003, the X-ray position of the southeast component of the dAGN (i.e., 2206$+$0003\,SE) shows an offset of 0.26\arcsec\ from the VLA 6\,GHz source. We correct for this astrometric offset in the ASOL file and the event-2 file with \texttt{wcs\_update}. The other three observations show no significant astrometric offset wherever there are sufficient number of X-ray sources to carry out a positional comparison between the X-ray sources from \texttt{wavdetect} and the radio sources from our VLA observations and the near-IR sources from UKIRT IR Deep Sky Survey \citep[UKIDSS;][]{Warren07}. 

Figure\,\ref{fig:xrayimgs} shows the full-band (0.3-8 keV) X-ray images of the targets, along with their near-IR images, and radio maps. The first impression of the figure is that four of the eight nuclei are clearly detected (0051$+$0020\,NE, 0051$+$0020\,SW, 2206$+$0003\,SE, and 2300$-$0005\,NW), two are probably detected (2206$+$0003\,NW and 2300$-$0005\,SE), and two are clearly undetected (2232$+$0012\,NW and 2232$+$0012\,SE). We present more quantitative results from aperture photometry in the next section.

\section{X-ray Data Analysis} \label{sec:analysis}

\begin{deluxetable*}{lccccccccc}
\tabletypesize{\small} \tablewidth{0pt}
\tablecaption{X-ray Observations and Photometry 
\label{tab:xraydata}}
\tablehead{ 
\colhead{Object} & \colhead{$z$} & \colhead{Exptime} & \colhead{Full} & \colhead{Soft} & \colhead{Hard} & \colhead{HR} & \colhead{$\log N_{\rm H}$} & \colhead{$\log F_{\rm 0.3-8keV}^{\rm abs}$} & \colhead{$\log F_{\rm 0.3-8keV}^{\rm unabs}$} \\ 
\colhead{} & \colhead{} & \colhead{(ks)} & \colhead{src (bkg)} & \colhead{src (bkg)} & \colhead{src (bkg)} & \colhead{} & \colhead{(cm$^{-2}$)} & \colhead{(erg~s$^{-1}$~cm$^{-2}$)} & \colhead{(erg~s$^{-1}$~cm$^{-2}$)} \\ 
\colhead{(1)} & \colhead{(2)} & \colhead{(3)} & \colhead{(4)} & \colhead{(5)} & \colhead{(6)} & \colhead{(7)} &  \colhead{(8)} & \colhead{(9)} & \colhead{(10)}
}
\startdata 
0051$+$0020\,SW&0.11253&28.7&30 (0.3)&19 (0.1)&11 (0.2)&$-0.28_{-0.17}^{+0.18}$&$21.2_{-1.2}^{+0.5}$&$-13.9_{-0.1}^{+0.1}$&$-13.8_{-0.2}^{+0.2}$\\
0051$+$0020\,NE&0.11257&28.7& 6 (0.3)& 3 (0.1)& 3 (0.2)&$-0.02_{-0.40}^{+0.40}$&$21.9_{-1.9}^{+0.5}$&$-14.6_{-0.2}^{+0.2}$&$-14.3_{-0.4}^{+0.4}$\\
\\
2206$+$0003\,NW&0.04656&15.9& 3 (0.2)& 2 (0.1)& 1 (0.1)&$-0.63_{-0.26}^{+0.53}$&                  \nod &$-14.6_{-0.3}^{+0.3}$&$-14.6_{-0.3}^{+0.3}$\\
2206$+$0003\,SE&0.04640&15.9&72 (0.2)& 7 (0.1)&65 (0.1)&$+0.82_{-0.08}^{+0.06}$&$22.7_{-0.1}^{+0.1}$&$-13.0_{-0.1}^{+0.1}$&$-12.4_{-0.1}^{+0.1}$\\
\\
2232$+$0012\,NW&0.22128&16.9& 0 (0.2)& 0 (0.1)& 0 (0.1)&                  \nod &                  \nod &               $<-14.4$& $<-14.4$\\
2232$+$0012\,SE&0.22187&16.9& 1 (0.2)& 0 (0.1)& 1 (0.1)&                  \nod &                  \nod &               $<-14.3$& $<-14.2$\\
\\
2300$-$0005\,NW&0.17971&34.7&12 (0.6)&10 (0.3)& 2 (0.3)&$-0.78_{-0.17}^{+0.23}$&                  \nod &$-14.4_{-0.1}^{+0.1}$&$-14.3_{-0.1}^{+0.1}$\\
2300$-$0005\,SE&0.17981&34.7& 6 (0.6)& 4 (0.3)& 2 (0.3)&$-0.47_{-0.35}^{+0.42}$&                  \nod &$-14.7_{-0.2}^{+0.2}$&$-14.7_{-0.2}^{+0.2}$\\

\enddata
\tablecomments{ 
Rows are grouped in sets of two that each include a pair, and sources are sorted in ascending R.A.
(1) Object name;
(2) spectroscopic redshift;
(3) ACIS-S exposure time in kilosecond;
(4$-$6) source and background counts inside a 2.5\arcsec-diameter circular aperture centered on the radio position for full band ($F$, 0.3$-$8\,keV), soft band ($S$, 0.3$-$2.0\,keV), and hard band ($H$, 2.0--8.0\,keV), respectively;
(7) hardness ratio (HR) determined using Bayesian estimation method \citep[BEHR;][]{Park06}, which is defined as HR\,$=(H-S)/(H+S)$; quoted are the mode and the 15.8-and-84.1-percentiles;
(8) intrinsic hydrogen column density inferred from HR by fixing the photon index at $\Gamma = 1.9$. Three of the six detected sources do not have $N_{\rm H}$ measurements because their nominal HRs are softer than the assumed power-law;
(9$-$10) Absorbed and unabsorbed power-law model flux at the observed-frame full band. The absorbed flux includes both intrinsic absorption and Galactic absorption. Detections are quoted with the 1$\sigma$ uncertainty, and non-detections are listed as upper limits at 99\% confidence level. The quoted error of the unabsorbed flux accounts for the uncertainties in both the count rate and the HR-derived intrinsic column density.
}
\end{deluxetable*}

For source photometry, we adopt a 2.5\,pixel-radius circular aperture centered at the VLA 6\,GHz position (the native CCD pixel size of ACIS-S is 0.4920\arcsec). This corresponds to aperture radii of 2.5 kpc, 1.1 kpc, 4.4 kpc, and 3.7 kpc for the sources in 0051$+$0020, 2206$+$0003, 2232$+$0012, and 2300$-$0005, respectively. The source aperture encloses 92\%/88\% of the PSF at an effective energy of 1.5/3.6\,keV from \texttt{arfcorr} and its diameter of 2.46\arcsec\ is slightly less than the angular separation of the most closely separated pair in our sample (thus avoiding aperture overlapping). The aperture is also large enough to fully enclose the 90\% ACIS-S positional uncertainty radius of $\sim$0.7\arcsec\ (\url{http://cxc.harvard.edu/cal/ASPECT/celmon/}). To estimate the background contribution to the source counts, we define a background region using an annulus centered at the middle point of each pair with an inner and an outer radius of 20 pixels and 60 pixels, respectively. We exclude sources detected by \texttt{wavdetect} inside the background annulus with a 10\,pixel-radius circular region around each source. Counts from the source and background regions are measured with \texttt{srcflux} for three energy bands: soft ($S$; 0.3$-$2\,keV, $E_{\rm eff} = 1.5$\,keV), hard ($H$; 2$-$8\,keV, $E_{\rm eff} = 4.3$\,keV), and full ($F$; 0.3$-$8\,keV, $E_{\rm eff} = 3.6$\,keV), where the effective energy ($E_{\rm eff}$) is defined as the effective-area-weighted mean energy of each band. The full-band source counts range between 0 and 72, whereas the background is expected to contribute between 0.2 and 0.6 counts in the source aperture. The significance of the X-ray detections can be estimated using Poisson statistics. For the six AGN in 0051$+$0020, 2206$+$0003, and 2300$-$0005, the probability of the background counts constituting the observed counts in the aperture is less than 0.1\% in the 0.3$-$8\,keV band, which is equivalent to a confidence level of $>$99.9\% for a detection of the source. We thus consider these six sources as robust detections. Both AGN in 2232$+$0012 are undetected, so we estimate the 3$\sigma$ (99\%) upper limits of the count rates using the background-marginalized Bayesian algorithm built into \texttt{aprates} and convert them to flux upper limits assuming an unabsorbed power law model with $\Gamma = 1.9$, which is appropriate for AGN \citep[e.g.,][]{Just07,She17}. The statistical results are consistent with the visual impression of Fig.~\ref{fig:xrayimgs}. The total counts and the expected background counts in the three energy bands are listed in Table\,\ref{tab:xraydata}, along with a few derived parameters discussed below.

We estimate the hardness ratio, HR\,$\equiv (H-S)/(H+S)$, of the six detections using the Bayesian estimation method of \citet{Park06}. Assuming an intrinsic power-law spectrum with a fixed photon index, we can use the estimated HR to infer the intervening total Hydrogen column density in the host galaxy ($N_{\rm H}$). We again adopt the canonical power-law photon index of $\Gamma = 1.9$. For each target, we use the \texttt{modelflux} task to calculate the relation between HR and $N_{\rm H}$ for a redshifted power-law model with $\Gamma = 1.9$ absorbed intrinsically by the torus or the host galaxy and subsequently by $N_{\rm H}^{\rm Gal}$ from the Milky Way (\texttt{xszpowerlw * xszphabs * xsphabs}). The photoelectric absorption cross-sections \citep{Balucinska-Church92} assume the standard Solar abundance set from \citet{Anders89}. The HR$-$$N_{\rm H}$ relation varies among the sources because of the differences in redshift, the Galactic column density ($N_{\rm H}^{\rm Gal}$), and the Auxiliary Response Function (ARF) and Redistribution Matrix Function (RMF) specific to the \chandra\ observation. The Galactic column densities from the NRAO catalog \citep{Dickey90} range between $2.66\times10^{20}$\,cm$^{-2}$ and $5.21\times10^{20}$\,cm$^{-2}$. 

For three of the six detections (both sources in 0051$+$0020 and 2206$+$0003\,SE) that have HRs greater than the unabsorbed power-law model (i.e., ${\rm HR} > -0.4$), we can apply an HR-based method to self-consistently convert the count rate to absorbed and unabsorbed fluxes. We first use the HR$-$$N_{\rm H}$-relation and the measured HR to estimate $N_{\rm H}$, obtaining values for these three sources of $1.6\times10^{21}$\,cm$^{-2} \lesssim N_{\rm H} \lesssim 5\times10^{22}$\,cm$^{-2}$. We then scale the PSF-corrected count rate to the absorbed and the unabsorbed flux in each energy band using the scaling factors from \texttt{modelflux} for the HR-derived $N_{\rm H}$. The X-ray emission from these three sources appears spatially concentrated on the radio positions in Fig.\,\ref{fig:xrayimgs} (as expected for point sources) and their HRs are consistent with nuclear emission typically observed in AGN. Therefore, we conclude that the X-ray emission in 0051$+$0020 and 2206$+$0003\,SE are powered by BH accretion and one of the three AGN are moderately obscured ($N_{\rm H} > 10^{22}$\,cm$^{-2}$). We note that although AGN are observed to have a range of photon indices, our choice of $\Gamma = 1.9$ does not significantly impact our results given the HR values for our sample. Varying the photon index by $\Delta \Gamma$ = 0.5 in either direction yields $\Delta$log($N_{\rm H}\, /\,{\rm cm}^{-2}$) = 0.4, 0.12, and 0.07 dex for HR = 0, 0.5, and 0.8, respectively. The HRs of the other three detections (2206$+$0003\,NW, and both sources in 2300$-$0005) are lower than the HR of an unabsorbed power-law model (${\rm HR} < -0.4$), so we calculate their fluxes assuming negligible intrinsic absorption.  

Spectral fitting is only possible for 2206$+$0003\,SE, which has 72 counts. We group the extracted spectrum to a minimum of five counts per bin and fit the grouped spectrum with the same absorbed power-law model using \texttt{Sherpa}. We assume the same absorbed power-law model as above and allow $\Gamma$ to vary between 1.4 and 2.4. We adopt Cash statistics because of the low counts. We find a best-fit column density of $\log(N_{\rm H}/{\rm cm}^{-2}) = 23.0\pm0.2$, which is consistent with the HR-derived value within 1.5$\sigma$.

\section{Results} \label{sec:results}

\begin{deluxetable*}{lcccccccc}
\tabletypesize{\small} \tablewidth{0pt}
\tablecaption{Multi-Wavelength Properties
\label{tab:luminosity}}
\tablehead{ 
\colhead{Object} & \colhead{$\sigma_{\star}$} & \colhead{$\log M_{\rm BH}$} & \colhead{$\log L_{\rm H\alpha}$} & \colhead{$\log L_{\rm [OIII]}$} & \colhead{$\log L_{\rm 5GHz}$} & \colhead{$\log L_{\rm 12\mu m}$} & \colhead{$\log L_{\rm X}$} & \colhead{$\log L_{\rm X}^{\rm XRB}$}\\
\colhead{} & \colhead{(km~s$^{-1}$)} & \colhead{(\msun)} & \colhead{(erg~s$^{-1}$)} & \colhead{(erg~s$^{-1}$)} & \colhead{(erg~s$^{-1}$)} & \colhead{(erg~s$^{-1}$)} & \colhead{(erg~s$^{-1}$)} & \colhead{(erg~s$^{-1}$)} \\
\colhead{(1)} & \colhead{(2)} & \colhead{(3)} & \colhead{(4)} & \colhead{(5)} & \colhead{(6)} & \colhead{(7)} &  \colhead{(8)} & \colhead{(9)}
}
\startdata 
0051$+$0020\,SW&$140\pm12$&$7.8\pm0.3$&$41.80\pm0.02$&$41.33\pm0.11$&$38.73\pm0.02$&$43.88\pm0.02$&$41.5_{-0.2}^{+0.2}$&$40.0$\\
0051$+$0020\,NE&$166\pm 9$&$8.1\pm0.3$&$42.36\pm0.01$&$41.37\pm0.18$&$39.22\pm0.01$&          \nod&$41.0_{-0.4}^{+0.4}$&$40.4$\\
\\
2206$+$0003\,NW&$114\pm 9$&$7.4\pm0.3$&$41.23\pm0.01$&$40.63\pm0.06$&$37.37\pm0.10$&$42.77\pm0.02$&$39.9_{-0.3}^{+0.3}$&$39.5$\\
2206$+$0003\,SE&$170\pm 6$&$8.2\pm0.3$&$41.55\pm0.01$&$41.20\pm0.03$&$38.22\pm0.01$&          \nod&$42.0_{-0.1}^{+0.1}$&$39.7$\\
\\
2232$+$0012\,NW&$244\pm36$&$8.9\pm0.4$&$41.18\pm0.04$&$40.47\pm0.08$&$39.03\pm0.04$&$44.17\pm0.02$&               $<$41.5&$39.7$\\
2232$+$0012\,SE&$210\pm61$&$8.6\pm0.6$&$41.51\pm0.04$&$41.15\pm0.08$&$39.28\pm0.02$&          \nod&               $<$41.7&$39.9$\\
\\
2300$-$0005\,NW&$284\pm13$&$9.2\pm0.3$&      $<$40.50&      $<$40.05&$39.40\pm0.01$&$43.28\pm0.17$&$41.4_{-0.1}^{+0.1}$&$<$40.4\\
2300$-$0005\,SE&$324\pm11$&$9.4\pm0.3$&      $<$40.62&      $<$40.12&$38.52\pm0.09$&          \nod&$41.0_{-0.2}^{+0.2}$&$<$40.5
\enddata
\tablecomments{ 
(1) Object name;
(2) stellar velocity dispersion from fitting SDSS and Keck/LRIS spectra with stellar population synthesis models \citepalias{Fu15b};
(3) black hole mass inferred from the $M_{\rm BH}-\sigma_\star$ relation of \citet{Kormendy13}, which has an intrinsic scatter of $\sim$0.29\,dex;
(4-5) H$\alpha$ and [O\,{\sc iii}]\,$\lambda$5007 line luminosities, corrected for reddening and aperture-loss;
(6) rest-frame 5\,GHz luminosity ($\nu L_{\nu}$) computed from the VLA flux density at 6\,GHz \citepalias{Fu15b};
(7) rest-frame 12\,\um\ luminosity from AllWISE photometry (listed is the total luminosity for each pair because the components are blended in WISE images);
(8) unabsorbed rest-frame 2-10\,keV luminosity $K$-corrected from the unabsorbed model flux ($F_{\rm 0.3-8keV}^{\rm unabs}$) in Table\,\ref{tab:xraydata};
(9) expected X-ray luminosity from X-ray binaries.
}
\end{deluxetable*}

Our \chandra\ observations provide the X-ray fluxes and the HRs of our dAGN, and for several cases, the column densities of the intervening gas (Table\,\ref{tab:xraydata}). To estimate the unabsorbed X-ray luminosity at rest-frame 2$-$10\,keV ($L_{\rm X} \equiv L_{\rm 2-10 keV}$), we use the $K$-correction factor of the assumed power-law model to convert the unabsorbed 0.3$-$8\,keV flux to the rest-frame flux in 2$-$10\,keV. For the range of column densities observed in our sources ($N_{\rm H} \leq 10^{23}$\,cm$^{-2}$), the 2$-$10\,keV luminosity is almost unaffected by absorption: photoelectric absorption by intervening gas with $N_{\rm H} = 10^{23}$\,cm$^{-2}$ only decreases the 2-10\,keV luminosity by 0.28\,dex. In Table\,\ref{tab:luminosity}, we list the X-ray luminosities of the dAGN along with their properties observed at other wavelengths.

In the following sub-sections, we perform a series of comparisons between the dAGN and the general AGN population to assess whether the dAGN behave differently. We begin in \S~\ref{sec:Contamination} by establishing the possible non-AGN sources of X-ray emission in our sample. We then compare the general AGN properties such as radio-to-X-ray ratio, Eddington ratio, and the BH fundamental plane in \S~\ref{sec:Lx}. In \S~\ref{sec:HR} we analyze the distribution of X-ray hardness ratios to test whether the AGN in mergers are more obscured than isolated AGN, as expected from tidally induced inflows. Finally, in \S~\ref{sec:deficit} we check if the X-ray luminosities of dAGN are lower than expected from their mid-IR and [O\,{\sc iii}] luminosities and discuss the role of star formation in the dAGN. 

\subsection{Accounting for X-ray Contamination} \label{sec:Contamination}

The AGN in our sample have unabsorbed rest-frame 2-10\,keV luminosities in the range of $39.9 <$ log$\,L_{\rm X}/{\rm erg~s}^{-1} < 42.0$, which is in the regime between nearby LLAGN and Seyferts \citep{Brusa07}. In this regime, galactic sources of X-rays might be contributing significantly to the observed X-ray luminosity. Of particular concern is the contribution to \lx\ from X-ray binaries (XRBs), whose luminosity range can potentially overlap with LLAGN. Given the composite status and young stellar ages uncovered through BPT diagnostics in \citetalias{Fu15a} and spectral fitting in \,\citetalias{Fu15b}, respectively, the effects of XRBs should be taken into consideration. We first estimate the SFR using the extinction- and aperture-loss-corrected H$\alpha$ luminosities for our sample in the $L_{\rm H\alpha}$-SFR relation from \citet{Murphy11}:
\begin{equation}
\log({\rm SFR/M_\odot\ yr^{-1}}) = \log(L^{\rm SF}_{\rm H\alpha}/{\rm erg\ s^{-1}}) - 41.3.
\label{eq:lsfHa}
\end{equation}
These H$\alpha$-based SFRs should be considered as upper limits because a significant fraction of the H$\alpha$ line may come from AGN-photoionized gas. We then use this expected SFR to calculate the expected \lx\ due to XRBs using the calibration from \citet{Lehmer10}:
\begin{equation}
L^{\rm XRB}_{\rm X} = \alpha \times M_{*}+ \beta \times {\rm SFR ,}
\label{eq:XRB}
\end{equation}
where $\alpha= (9.05\pm0.37)\times10^{28} ({\rm erg\ s^{-1}})/$\msun\ and $\beta= (1.62\pm0.22)\times10^{39} {\rm erg\ s^{-1}}/$(\msunyr). The first term traces the low-mass XRBs and the second term traces the high-mass XRBs. We use the stellar masses of the galaxies obtained through stellar population synthesis modeling in \,\citetalias{Fu15b}. We list the predicted values of \lx\ due to XRBs in Table \,\ref{tab:luminosity}. For seven of the eight sources, the observed \lx\ are more than $\sim$0.6\,dex above the predicted \lx\ from XRBs. For 2206$+$0003\,NW, the difference is less than 0.4\,dex, indicating significant contribution from XRBs. This is also consistent with the low HR ($-$0.63) of the source. 

On the other hand, the low HRs of both sources in 2300$-$0005 is likely dominated by diffuse kpc-scale thermal plasma with $T \sim 10^7$\,K (i.e., $kT \sim 1$\,keV) in the host galaxies \citep[e.g.,][]{Donato04}. For reference, the intrinsic HR is about $-0.8$ for an \texttt{apec} thermal plasma model with $kT = 1$\,keV and $Z = 0.2\,Z_\odot$. Indeed, the X-ray emission of both sources appears spatially extended (e.g., simply compare 2300$-$0005 with 0051$+$0020 in Fig.\,\ref{fig:xrayimgs}), although their counts are too low to allow a robust measurement of the radial profiles.

To sum up, we find that our sample consists of three sources (both objects in 0051$+$0020 and 2206$+$0003\,SE) with definite AGN emission in X-ray, one source (2206$+$0003\,NW) with non-negligible contributions from XRBs, two sources (both objects in 2300$-$0005) with strong contributions from hot ISM, and two undetected sources (both objects in 2232$+$0012). For the latter five sources, the observed \lx\ should be considered as upper limits of the AGN X-ray luminosity.

\subsection{General AGN properties} \label{sec:Lx}

\begin{figure}
\includegraphics[width=8.5cm]{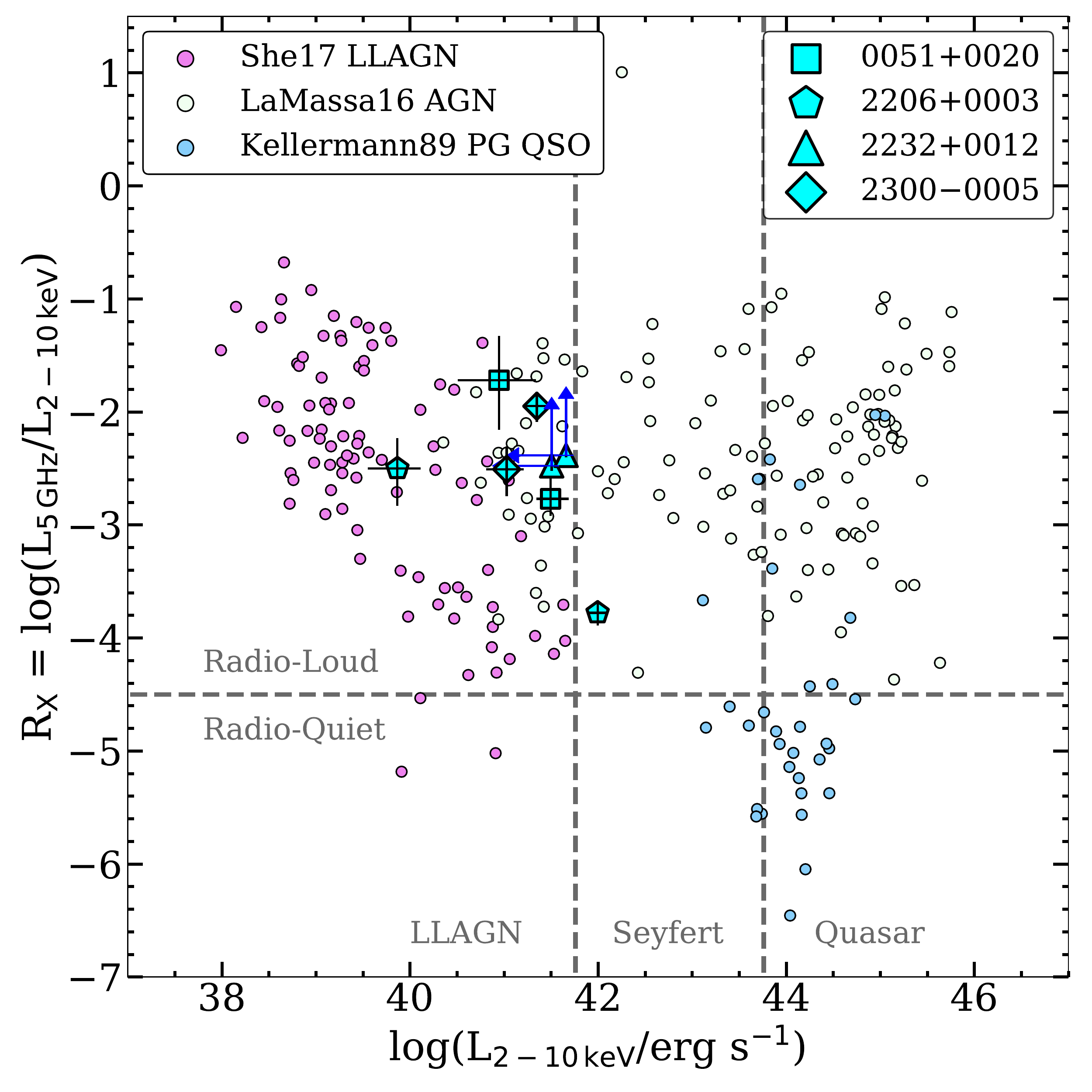}
\caption{Radio-to-X-ray luminosity ratio ($R_{\rm X}$) vs. unabsorbed rest-frame 2-10\,keV luminosity (\lx). The cyan points show our sample, where a pair is denoted by the same shape. The small color-filled circles show the comparison samples described in \S\,\ref{sec:Lx} and are labeled in the legend on the upper left. Horizontal arrows indicate X-ray upper-limits, and vertical arrows indicate lower limits in $R_{\rm X}$. The vertical dashed lines at $\log (L_{\rm X}/{\rm erg\,s}^{-1}) = 41.76$ and $43.76$ roughly divide AGN into three luminosity regimes: LLAGN, Seyfert, and Quasar \citep{Brusa07}. The horizontal dashed line separates radio-loud ($R_{\rm X} > -4.5$) and radio-quiet ($R_{\rm X} < -4.5$) AGN. The dAGN have X-ray luminosities in the LLAGN regime and show similar radio-to-X-ray ratios.}
\label{fig:Rx}
\end{figure}

We now compare the observed properties of our dAGN sample with those of AGN in isolated galaxies. In Fig.~\ref{fig:Rx}, we plot $R_{\rm X}$, the radio-to-X-ray luminosity ratio defined as $R_{\rm X} \equiv \log(L_{\rm 5GHz}/L_{\rm X})$, against $L_{\rm X}$. In this paper, the monochromatic radio luminosity is defined as $\nu L_\nu$ at rest-frame 5\,GHz.
To compare our dAGN with the general AGN populations, we show three comparison samples from the literature in Fig.\,\ref{fig:Rx}.
First, in the low luminosity regime ($\log L_{\rm X} \lesssim 42$\,\ergs), we adopt $L_{\rm X}$ from a sample of nearby ($< 50$ Mpc) LLAGN from \citet{She17} and cross-match the sources with the FIRST 1.4\,GHz catalog \citep{Helfand15} to obtain the radio luminosity. 
Second, in the high luminosity regime ($\log L_{\rm X} \gtrsim 42$\,\ergs), we plot the X-ray-selected AGN from the Stripe 82 X-ray Survey \citep{Lamassa16}. Similar to the LLAGN sample, a $K$-correction was applied to the 1.4\,GHz flux density assuming $\alpha = 0.7$. We also $K$-correct the observed luminosities in 0.5-7\,keV (\chandra) and 0.5-10\,keV (\xmm) to rest-frame 2$-$10 keV assuming $\Gamma = 1.9$.
Lastly, we include a subset of Palomar-Green (PG) QSOs, for which we compiled 5\,GHz fluxes from \citet{Kellermann89} and \xmm\ X-ray luminosities from \citet{Piconcelli05}. 
The breadth of the parameter space covered by these samples allows for a comprehensive comparison. This figure shows that the dAGN follow the same radio-X-ray scaling relation established by the general population of AGN which are mostly hosted by isolated galaxies. 

Note that even though we show a horizontal line at $R_{\rm X} = -4.5$ to separate between radio-loud and radio-quiet AGN, $R_{\rm X}$ itself is not a robust indicator for accretion. In fact, star-forming galaxies have similar radio-to-X-ray ratios as the LLAGN in Fig.\,\ref{fig:Rx}. One obtains $R_{\rm X} = -2.2$ for pure star-forming galaxies using the $L_{\rm 1.4\ GHz}$-SFR relation scaled to 5 GHz from \citet{Murphy11} and the \lx-SFR relation from \citet{Ranalli03}:
\begin{equation}
\log(L^{\rm SF}_{\rm X} /{\rm erg\ s^{-1}}) = \log({\rm SFR/M_\odot\ yr^{-1}}) + 39.7,
\label{eq:Lsf1}
\end{equation}
which constitutes an upper limit on the \lx\ from star-formation related effects. On the other hand, $R_{\rm X}$ is directly related to the radio-to-X-ray spectral index between 5\,GHz and 2\,keV ($4.84\times10^{17}$\,Hz):
\begin{equation}
 \alpha_{\rm RX} = (R_{\rm X}+8.242)/8 ~~~{\rm for}~~~ \Gamma = 1.9.
\end{equation}
For the average value of $R_{\rm X} = -2.5$ for our dAGN, $\alpha_{\rm RX}$ equals 0.72, consistent with a single synchrotron spectrum extending from X-ray to radio \citep[e.g.,][]{Hardcastle01}. Nonetheless, this crude SED slope measurement is inadequate to rule out models that predict separate origins of the X-ray and the radio emission. 

\begin{figure}
\includegraphics[width=8.5cm, height=8.5cm]{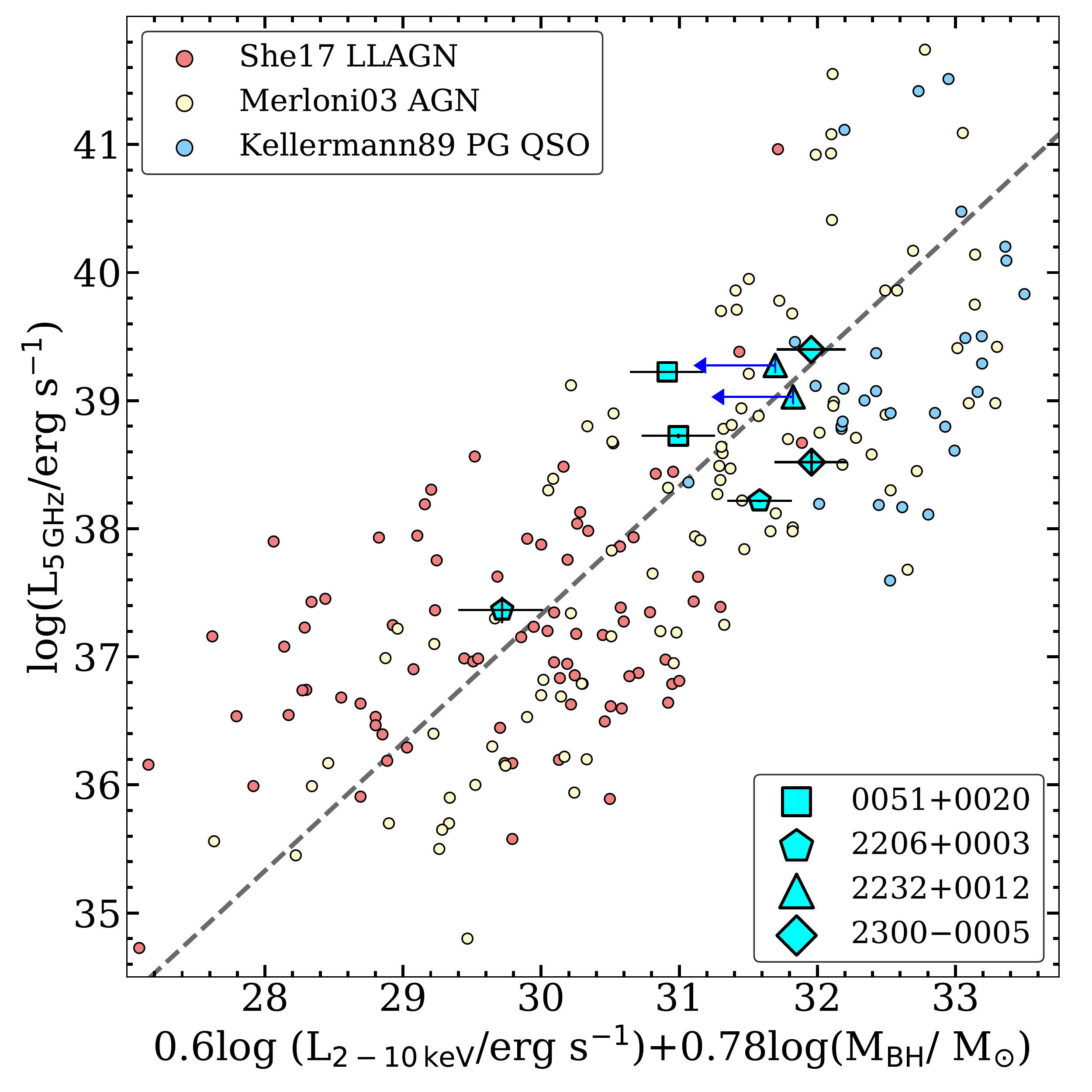}
\caption{The black hole fundamental plane \citep{Merloni03}. The dashed line follows the empirical relation given in Eq.~\ref{eq:BHFP}. The cyan points for our sample follow the same convention as in Fig.~\ref{fig:Rx}. The arrows indicate X-ray upper-limits. The smaller data points show the comparison samples described in \S\,\ref{sec:Lx}. Our dAGN follow well the fundamental plane relation with no systematic offsets.}
\label{fig:BHFP}
\end{figure}

The radio$-$X-ray correlation is closely related to the BH fundamental plane, which is established by both stellar-mass BHs in the Galaxy and the SMBHs in AGN and QSOs. The three-parameter empirical correlation is best-fit by the following power-law \citep{Merloni03}:
\begin{multline}
\log(L_{\rm 5GHz}/{\rm erg\,s}^{-1}) =  \\
0.6 \log(L_{\rm X}/{\rm erg\,s}^{-1}) + 0.78 \log(M_{\rm BH}/\rm M_\odot) + 7.33,
\label{eq:BHFP}
\end{multline}
where $L_{\rm 5GHz}$, $L_{\rm X}$, and $M_{\rm BH}$ are the rest-frame 5\,GHz luminosity of the radio core, rest-frame 2$-$10\,keV luminosity, and BH mass, respectively. Although the relation holds over an impressive range of BH mass and luminosity, it has a rather large scatter ($\sim$1\,dex) and both the disk-jet model \citep{Merloni03} and the jet-dominated model \citep{Falcke04} can explain the same correlation. 

With the X-ray luminosities from \chandra\ and the BH masses from the $M_{\rm BH}-\sigma_\star$ relation of nearby classical bulges and elliptical galaxies \citep{Kormendy13}, we can now place the dAGN on the fundamental plane in Fig.\,\ref{fig:BHFP}. Similar to Fig.\,\ref{fig:Rx}, we also plot the following comparison samples: (1) the AGN from the original \citet{Merloni03} sample, (2) LLAGN with \chandra/ACIS X-ray luminosity and BH mass from the $\rm M_{\rm BH}-\sigma_\star$ relation \citep{She17} and 1.4\,GHz radio flux from the FIRST survey \citep{Helfand15}, and (3) PG QSOs with 5\,GHz radio flux from \citet{Kellermann89}, \xmm\ X-ray luminosity from \citet{Piconcelli05}, and H$\beta$-based virial BH mass from \citet{Lani17}. It is clear from Fig.~\ref{fig:BHFP} that the dAGN follow the same fundamental plane relation as the other AGN samples. They are distributed well within the scatter of the relation, with no systematic deviation from the best-fit relation. 

Lastly, multiplying \lx\ by $\sim$16 to estimate the bolometric luminosity \citep{Ho08} and using the $\sigma_\star$-derived BH mass to estimate the Eddington luminosity for ionized hydrogen, we find that the dAGN have Eddington ratios between $-5 < \log L_{\rm bol}/L_{\rm Edd} < -3$, which again are similar to LLAGN. This last piece of evidence argues against a disk-corona model for the origin of nuclear X-rays in our dAGN.

In summary, we have compared the dAGN with other low-redshift AGN in terms of general AGN properties, such as radio-to-X-ray ratio, BH fundamental plane, and Eddington ratio. We find that the dAGN are similar to nearby LLAGN and Seyferts, suggesting that being involved in close galactic encounters does not affect these general AGN properties.

\subsection{Distribution of X-ray Hardness Ratios} \label{sec:HR}

Simulations suggest high gas column density in merging galaxies due to the central concentration of gas as the tidal torques funnel the ISM into the central kpc \citep{Hernquist89}. The merger-driven model of quasars predicts a wide distribution of column densities ($N_{\rm H} = 10^{21} - 10^{25}$\,cm$^{-2}$) with a peak around $10^{23}$\,cm$^{-2}$ \citep{Hopkins06b}. Our \chandra\ observations offer a way to test this prediction. As described in \S~\ref{sec:analysis}, we have obtained HR measurements for six sources in our dAGN sample. By assuming an unabsorbed photon index of the intrinsic AGN emission, the observed HRs can be used as a proxy for the column density, and we find that only one of the six sources are obscured (i.e., $N_{\rm H} > 10^{22}$\,cm$^{-2}$; see Table\,\ref{tab:xraydata}). In this subsection, we examine whether the distribution of the HR values is different from that of the general LLAGN. 

To construct the control sample, we select 68 AGN with X-ray luminosities between $39.8 < \log L_{\rm X}/{\rm erg\,s^{-1}} < 42.0$ from the catalog of 314 nearby LLAGN observed by \chandra\ \citep{She17}. The luminosity range matches that of the dAGN, so it prevents potential biases of the HRs due to any possible correlation between $N_{\rm H}$ and $L_{\rm X}$. The control data set is an archival compilation obtained over many \chandra\ cycles, wherein the sensitivity of the detectors in soft X-ray is known to degrade over time due to increased contamination. Conversely, the effective area decreases for off-axis observations, and the effective area decreases faster in the hard energy band than the soft band. In addition, different detector arrays (ACIS-I vs. ACIS-S) were used in the archival observations. For a fair comparison with the HR values of the dAGN, we institute correction factors to convert the reported HR values of the control sample to reflect the characteristics of ACIS-S in Cycle\,18. 
Correction factors are computed utilizing the \texttt{modelflux} task with the ARF and RMF files of each source (kindly provided by Rui She), which encapsulate all of the above effects.

\begin{figure}
\includegraphics[width=8.5cm]{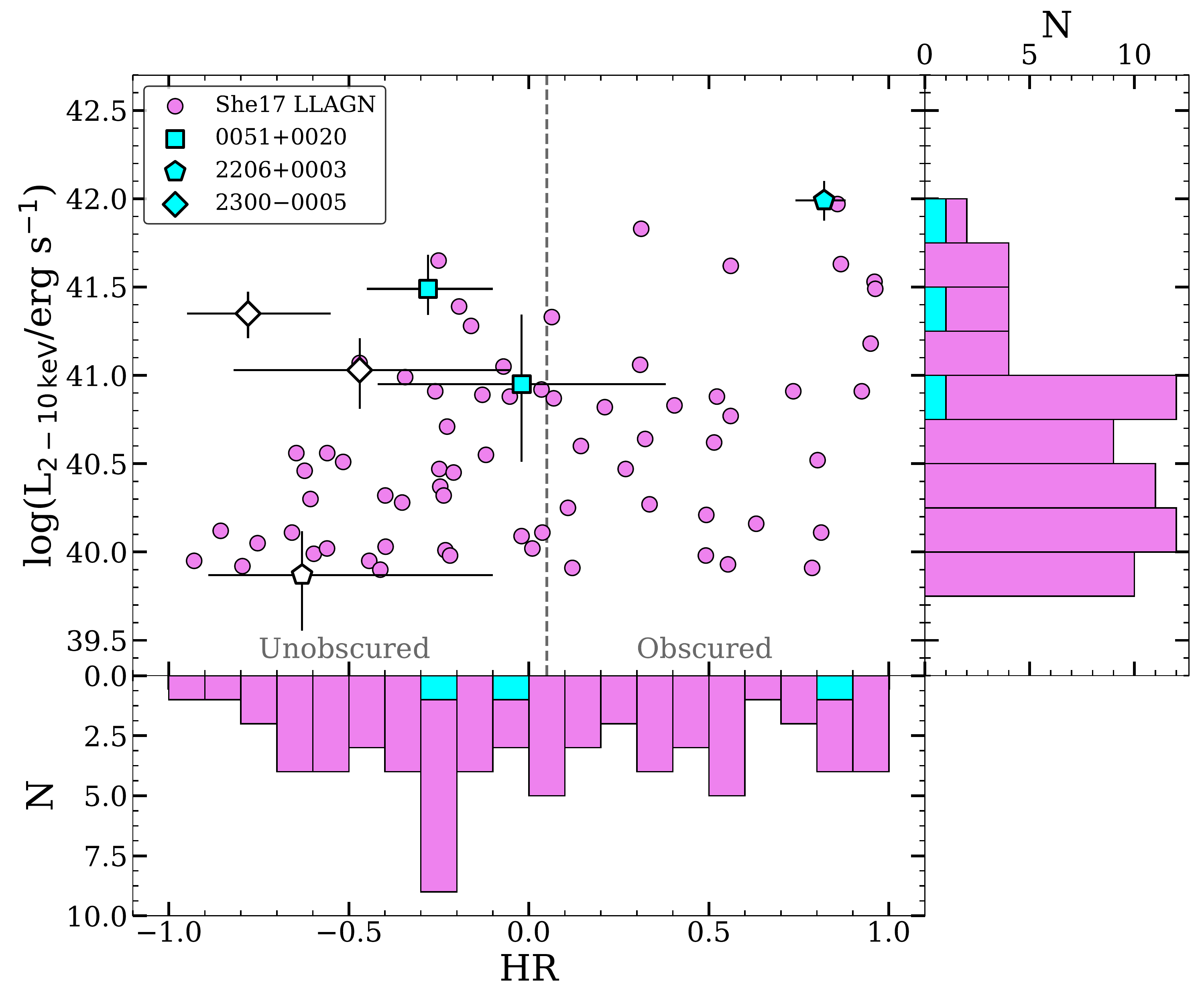}
\caption{X-ray luminosity vs. hardness ratio. We show AGN-dominated components in our sample as {\it filled} shapes and histograms. {\it Open} shapes indicate the objects in our sample which have measured HR values, but are likely dominated by non-nuclear sources of soft X-ray emission as discussed in \S~\ref{sec:Contamination}. We also show the control sample of nearby LLAGN from \citet{She17} ({\it pink} circles and histograms). The HR distribution of the dAGN is roughly consistent with that of the control sample. The vertical dashed line separates unobscured and obscured AGN.}
\label{fig:HR_compare}
\end{figure}

In Fig.~\ref{fig:HR_compare}, we compare our dAGN sample and the control sample of LLAGN in the plane of X-ray luminosity versus HR. At first glance, the dAGN appear to be systematically softer X-ray sources than the control sample, contrary to the expectation that merger-driven inflows produce higher X-ray obscurations in the host galaxies, and thus higher HR values. Note that in the following comparison, we do not include either source in 2300$-$0005 or 2206$+$0003\,NW, because their diffuse soft X-ray emission may originate from an interstellar thermal plasma or from XRBs, respectively, instead of nuclear accretion (\S~\ref{sec:Contamination}). Between $39.8 < \log L_{\rm X}/{\rm erg\,s^{-1}} < 42.0$, 30 out of the 68 LLAGN in \citet{She17} have HR\,$> 0.05$, which corresponds to $N_{\rm H} > 10^{22}$\,\cmsq\ for $\Gamma = 1.9$. This fraction of obscured AGN is consistent with the previously determined distribution of column densities as a function of X-ray luminosity from \citet{Ueda03}. In contrast, only one out of the  three dAGN components have HR\,$> 0.05$. Limited by the small sample size, the difference between the two obscured  AGN fractions is within 1$\sigma$ --- $44\pm6$\% vs. $33^{+28}_{-15}$\% --- where the 1$\sigma$ confidence intervals are estimated using the Bayesian approach for binomial population proportions \citep{Cameron11}. Therefore, the X-ray HRs of these dAGN are normal for their luminosity, lending no evidence that X-ray emission is more obscured in these galaxy mergers than in isolated AGN, even when the projected separations are less than 10\,kpc. Given that only six of the individual components in our sample have measurable HR values, and three of these may be dominated by non-nuclear soft X-ray emission, a larger and deeper sample is needed to confirm this result.

\subsection{Elevated Obscuration or Contamination from Star Formation?} \label{sec:deficit}

Previous work argues that dAGN show higher degrees of obscuration by comparing the X-ray luminosity with the mid-IR (12\,\um) luminosity \citep{Satyapal17}. In this case, the observed X-ray luminosities of dAGN are lower than expected from the other AGN luminosity tracer. In this subsection, we apply the same diagnostic tools on the radio-selected dAGN and discuss their limitations. 

\begin{figure}
\includegraphics[width=8.5cm]{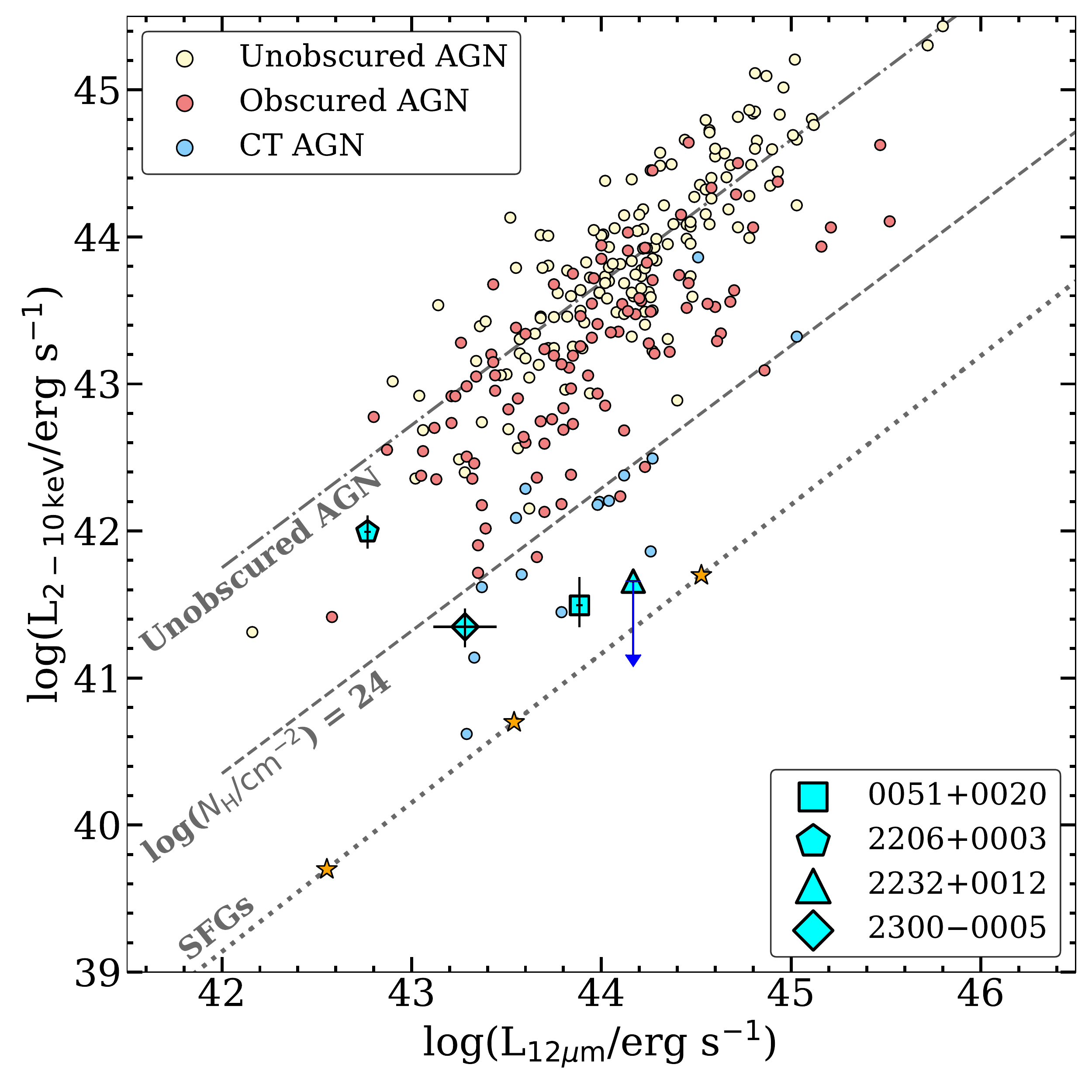}
\caption{Luminosity relation sensitive to X-ray obscuration ($L_{\rm X})$ vs. rest-frame 12\,\um\ luminosity from WISE (\lmir). The large cyan data points with error bars show our dAGN and the small circles are BAT AGN from \citet{Ricci15}. The dot-dashed line indicates the relation for the unobscured AGN in the comparison sample. The dashed line shows the predicted relation for AGN obscured by a column density of $N_{\rm H} = 10^{\rm 24}$\,\cmsq\, calculated by \citet{Satyapal17} using a torus model that absorbs and reprocesses X-ray emission.  Due to the spatial resolution of WISE, the luminosities plotted for the dAGN are the sum of both components in each dual system. The dotted line shows the expected relation for star-forming galaxies, with stars marking SFRs at 1, 10, and 100\,\msunyr. Arrows denote upper limits. The dAGN show apparent X-ray deficits relative to the mid-IR luminosities, which may be due to enhanced \lmir\  from star formation rather than elevated X-ray obscuration.
}
\label{fig:LxLir}
\end{figure}

In Fig.~\ref{fig:LxLir} we plot X-ray luminosity against mid-IR luminosity, following \citet{Satyapal17}. For the dAGN, we interpolate the AllWISE photometry\footnote{\url{http://wise2.ipac.caltech.edu/docs/release/allwise/expsup/}} to obtain the rest-frame 12\,\um\ luminosities ($L_{\rm 12\mu m}$). Because our dAGN are spatially unresolved in WISE (FWHM $\sim$ 6\arcsec\ to 12\arcsec; \citealt{Wright10}), the 12\,\um\ luminosity is that of the whole dAGN system rather than that of individual components. To be consistent, we plot the combined X-ray luminosity for each dAGN. For comparison with the general AGN population, we show the hard X-ray selected nearby AGN from the 70 month Swift/BAT survey \citep{Baumgartner13, Ricci15}. The BAT AGN are denoted as unobscured ($N_{\rm H} < 10^{22}$\,\cmsq), obscured ($10^{\rm 22} < N_{\rm H} < 10^{\rm 24}$\,\cmsq), or Compton-thick ($N_{\rm H} > 10^{\rm 24}$\,\cmsq), based on column densities estimated from X-ray spectral analysis. As previously observed, unobscured BAT AGN follow a linear correlation between \lx\ and \lmir, while obscured and Compton-thick AGN systematically deviate from this correlation by showing an apparent deficit in X-ray luminosity. This is because the mid-IR photons emitted by the circumnuclear torus are less affected by obscuration than the X-ray photons. This plot thus provides a diagnostic tool to identify obscured AGN, albeit crude given the large scatter around the best-fit correlation. 

Similar to the previously X-ray observed dAGN compiled by \citet{Satyapal17}, the majority (3 out of 4) of the dAGN in our sample fall systematically below the \lx-\lmir\ correlation established by unobscured AGN and more closely resemble the Compton-thick AGN sample. Counterintuitively, the only obscured dAGN we identified based on the HR and the X-ray spectrum, 2206$+$0003, is in fact closest to the \lx-\lmir\ correlation. The apparent X-ray deficits for the remaining three dAGN imply high obscuring column densities. This is in tension with the low HR-derived column densities found in \S~\ref{sec:analysis}. To elucidate this matter, we show the expected location of star-forming galaxies as a dotted line in Fig.~\ref{fig:LxLir}, where we have used the \lmir-SFR relation from \citet{Donoso12}:
\begin{align}
\log(L^{\rm SF}_{12 \mu m}/{\rm erg\ s^{-1}}) &= 0.987~\log({\rm SFR/M_\odot\ yr^{-1}}) + 42.5,
\label{eq:Lsf3}
\end{align}
as well as the \lx-SFR relation given in Eq. \ref{eq:Lsf1}. Star-forming galaxies show a $\sim$2.3\,dex offset in \lx\ at any given \lmir. So an alternative way to explain the location of the dAGN is that their mid-IR emission is dominated by star formation. While our dAGN do fall below the trend for unobscured AGN, the mid-IR contribution from star formation (in 0051$+$0020, 2206$+$0003, and 2232$+$0012, which are classified as star-forming/AGN composites on the BPT diagram) and stars in the host galaxies (in 2300$-$0005, which has no active star formation) may offer alternative interpretations for the observed offsets other than elevated X-ray obscuration. There is a caveat to this approach for sources with significant non-nuclear soft X-ray emission (\S~\ref{sec:Contamination}). The observed HR of the source may artificially appear lower, potentially masking the absorbed signature of a highly obscured LLAGN. 

To assess whether our sources are dominated by AGN or star formation in the near IR, we check their WISE $W1 (3.4\mu m)- W2 (4.6\mu m)$ colors provided by \citet{Satyapal17}. Two of our dual systems are identified as AGN by WISE using a generous threshold of $W1 - W2 > 0.5$ (0051$+$0020 and 2232$+$0012). Assuming that the mid-IR contribution from star formation is unimportant in WISE-color-selected AGNs, the results in Fig.\,\ref{fig:LxLir} would indicate that the AGN X-ray emission in 0051$+$0020 and 2232$+$0012 are heavily obscured. This seems inconsistent with the soft X-ray HRs observed in 0051$+$0020, especially considering that XRBs are insignificant in this system (\S~\ref{sec:Contamination}). Unfortunately, to check the level of obscuration in 2232$+$0012, we would need much deeper X-ray observations. 

The [O\,{\sc iii}]$\lambda$5007 line emission from the kpc-scale extended AGN narrow-line region is less affected by nuclear obscuration than the nuclear X-ray emission. This makes the \lx-\loiii\,  ratio another diagnostic tool for obscuration similar to the \lx-\lmir\ diagnostic \citep{Panessa06}. \citet{Liu13a} found that the X-ray to [O\,{\sc iii}] luminosity ratios of double-peaked [O\,{\sc iii}]-selected dAGN are systematically lower than those of optically-selected single type 2 AGN. They suggest that the dAGN are systematically X-ray weak because of higher X-ray absorption columns and/or viewing angle bias from the double-peaked selection. We test whether a similar X-ray deficit is present in our sample of dAGN using the optical spectroscopy from \citetalias{Fu15b}. Comparing to the observed \lx$-$\loiii\ relation established by isolated AGN \citep{Panessa06, Trichas13}, our sample of dAGN show an average offset of $\sim$1.2 dex in \lx, suggesting higher levels of obscuration than the HR-derived column densities.  When considering only the three AGN-dominated components, the average offset is $\sim$1.5 dex, in comparison to the double-peaked [O\,{\sc iii}]-selected sample of dAGN presented in \citet{Liu13} which have an average X-ray deficiency of $\sim$2.5 dex. As well, the large scatter in the control sample of isolated AGN used to establish the observed \lx$-$\loiii\ relation encompasses the majority of our sample. So our radio-selected dAGN appear to be systematically less X-ray-weak than the double-peaked [O\,{\sc iii}]-selected dAGN. As with the \lx$-$\lmir\ comparison above, the contribution to the observed \loiii\ from star-formation may also explain the apparent offset as an enhancement of the [O\,{\sc iii}] luminosity as opposed to an X-ray deficiency from enhanced obscuration. As above, the caveat for artificially low HR values applies to the sources in our sample with contamination from non-nuclear soft X-rays.

\section{Summary and Conclusion} \label{sec:summary}

In this work, we have obtained \chandra\ ACIS-S observations of four kpc-scale radio-selected dAGN in Stripe 82. We detect X-ray emission from six of the eight dAGN components at $>$3$\sigma$ confidence level. Three X-ray components are consistent with low luminosity AGN. We find evidence for significant non-nuclear X-ray activity for the remaining three detected components. The X-ray and multi-wavelength properties of the dAGN are analyzed and compared with general AGN populations and previously studied dAGN. Our main results are summarized below:

\begin{enumerate}

\item The intrinsic rest-frame 2$-$10\,keV luminosities of our sample ranges between $39.9 < \log\,L_{\rm X}/{\rm erg~s}^{-1} < 42.0$, straddling the luminosity boundary between nearby LLAGN and Seyferts. With $7.4 < \log M_{\rm BH}/M_\odot < 9.4$ and $37.4 < \log L_{\rm 5 GHz}/{\rm erg~s}^{-1} < 39.4$, the components in the dAGN show low Eddington ratios and high radio-to-X-ray luminosity ratios similar to those of LLAGN, and they follow the same BH fundamental plane relation as the general AGN population. 

\item The X-ray hardness ratios indicate one obscured AGN (with $\log N_{\rm H}/{\rm cm}^{-2} = 22.7$) among the three AGN-dominated X-ray sources. X-ray spectral fitting of this source finds a similarly high column density. The fraction of obscured AGN ($33^{+28}_{-15}$\%) is comparable to that of nearby LLAGN in the same luminosity range ($44\pm6$\%). This result is at odds with simulation predictions of enhanced obscuration in advanced mergers due to large-scale gas inflows.

\item The dAGN show apparent X-ray deficiency with respect to the AGN luminosities inferred from the 12\,\um\ WISE photometry and the [O\,{\sc iii}] spectroscopy, similar to previously studied dAGN selected using other techniques. But it is important to account for the contribution from star-formation when interpreting the ``X-ray deficit,'' because the AGN luminosities inferred from mid-IR continuum and optical emission lines may have been significantly overestimated without subtracting the non-AGN components.

\end{enumerate}

Considering the multi-wavelength evidence, the radio-selected dAGN in Stripe 82 show properties similar to nearby LLAGN in terms of radio-to-X-ray ratio, the Eddington ratio, and BH mass. Although apparent X-ray deficiency is observed relative to mid-IR or [O\,{\sc iii}] luminosities in the dAGN, significant contribution to \lmir\ and \loiii\ from star formation (i.e., mid-IR and emission-line excess) provides a more natural explanation for the data. This alternative interpretation is also more consistent with (1) the agreement with the BH fundamental plane (\S~\ref{sec:Lx}), (2) the normal X-ray HR distribution (\S~\ref{sec:HR}), and (3) the AGN$-$star-forming composite emission-line ratios in the BPT diagram \citepalias[see][Fig.~6]{Fu15a}.

Despite our small sample, we find that these radio-selected dAGN show that being involved in close galactic encounters does not measurably alter their general AGN properties, nor does it increase their observed X-ray obscuring column density. These results seem to be in tension with the merger-driven scenario of AGN fueling, especially given that (1) these galaxies are in kpc-scale mergers, (2) at least two of the dAGN show spectacular tidal tails and shells that indicate a post-pericentric encounter, and (3) the number of observed dAGN exceeds by an order-of-magnitude the expectation from random pairing and the mean duty cycle of radio AGN in VLA-Stripe 82 \citepalias[3 observed vs. 0.3 expected $(13\times(551/22192))$; see][\S\,3.4]{Fu15b}. 

To reconcile these results, the most promising feeding mechanism to explain the low-level AGN activities in our dAGN seems to be stochastic-mode accretion due to tidally-triggered minor perturbations \citep[e.g.,][]{Hopkins06a}. In contrast to the conventional ``merger-driven accretion'' where the SMBH accretion is fueled by kpc-scale gas inflows and regulated by feedback \citep[e.g.,][]{Di-Matteo05,Hopkins05a}, the stochastic accretion naturally explains the similarities between the dAGN and LLAGN in apparently non-interacting hosts and the high frequency of correlated AGN (thus the higher-than-expected number of dAGN), but it evades the problem of enhanced obscuring column density due to large-scale tidal inflows (which is unobserved in our sample). Because a LLAGN with $L_{\rm bol} \lesssim 10^{10}$\,\lsun\ and a lifetime of 10\,Myr only requires a modest gas supply ($M \lesssim 6\times10^4/(\eta_{\rm rad}/0.1)$\,\msun, where $\eta_{\rm rad}$ is the radiative efficiency), large-scale gravitational torques from major mergers are not required to deliver the gas to the galactic nuclei. Instead, a sufficient amount of gas can be delivered to the nuclei even by a low-level cooling flow of the hot ISM \citep[e.g.,][]{Allen06}, and minor perturbations such as disk and bar instabilities \citep[e.g.,][]{Norman83,Jogee06}, magnetic breaking \citep[e.g.,][]{Krolik90}, $N$-body cloud-cloud or cloud-cluster interactions like in the Galactic center \citep[e.g.,][]{Genzel94}, and minor mergers \citep[e.g.,][]{Hernquist95}. Once the nuclear gas reservoir has been built up, the black hole can accrete gas through stochastic cloud collisions, and the accretion is subsequently self-regulated by feedback \citep[e.g.,][]{Yuan11,Bu19}. This stochastic accretion mode is distinct from the major-merger-induced accretion mode, and the former dominates the latter in the AGN population below the Seyfert/Quasar transition luminosity of $L_{\rm bol} = 10^{12}$\,\lsun\ \citep{Hopkins14a}. The low-level AGN in major mergers in our sample suggest that the stochastic mode has continued to operate in these systems even in the presence of large-scale tidal torques. The gas-rich major mergers may not have induced kpc-scale gas inflows (which would have obscured the X-ray), but they may have triggered minor perturbations in both nuclei simultaneously, which led to the high fraction of correlated LLAGN in mergers. This scenario represents a hybrid between the two above scenarios, namely merger-assisted stochastic accretion. This confluence of triggering mechanisms may be at play in some of the previously identified low luminosity dAGN systems as well \citep[e.g.,][]{Liu13a,Comerford15,Fu18}.

\acknowledgments

We thank the anonymous referee for thoughtful comments that helped improve the paper. We thank Philip Kaaret, Cornelia Lang, Joshua Steffen, and Dylan Par\'e for helpful discussions, and Hua Feng, Claudio Ricci, Shobita Satyapal, and Rui She for providing data from their previous analyses. The scientific results reported in this article are based on observations made by the Chandra X-ray Observatory. Support for this work was provided by the National Aeronautics and Space Administration (NASA) through Chandra Award Number GO7-18084X issued by the Chandra X-ray Center, which is operated by the Smithsonian Astrophysical Observatory for and on behalf of the National Aeronautics Space Administration under contract NAS8-03060. A.G. and H.F. acknowledge support from the National Science Foundation (NSF) grant AST-1614326. A.D.M. acknowledges support from NSF grant AST-1616168. S.G.D. acknowledges support from NSF grants AST-1413600 and AST-1518308, as well as NASA grant 16-ADAP16-0232. This research has made use of software provided by the Chandra X-ray Center (CXC) in the application packages CIAO, ChIPS, and Sherpa.

\bibliographystyle{aasjournal}

\end{document}